%% file: main.tex
\documentclass[11pt,letterpaper]{article}
 
\usepackage[margin=1in]{geometry} 
\usepackage[utf8]{inputenc}
\usepackage{microtype}
\usepackage{amsmath,amsthm,amssymb, color, tikz, mathtools}
\usepackage{tcolorbox}
\usepackage{mdframed}
% ...

\usetikzlibrary{positioning}
\usetikzlibrary{shapes}
\usetikzlibrary{decorations.pathreplacing}
\usepackage{enumerate, enumitem} 
\usepackage[pagebackref]{hyperref} 
\hypersetup{colorlinks,linkcolor={blue},citecolor={blue},urlcolor={blue}}

\newcommand{\ZZ}{\mathbb{Z}}

\newcommand{\norm}[1]{\left\lVert #1 \right\rVert}

\usepackage{bbm}

\newtheorem{theorem}{Theorem}[section]

\newtheorem{lemma}[theorem]{Lemma}

\newcommand{\ket}[1]{|#1\rangle}
\newcommand{\bra}[1]{\langle#1|}

\renewcommand{\epsilon}{\varepsilon}

\title{Improved Exact and Approximate Quantum Fourier Transforms
from Multiplication}
\author{Chandrima Kayal\footnote{Universit\'e Paris Cit\'e, CNRS, IRIF, Paris, France, Email:{\tt chandrimakayal2012@gmail.com}}
 \and Rajat Mittal\footnote{Indian Institute of Technology Kanpur. Email{\tt rmittal@cse.iitk.ac.in}}
 \and
 Manaswi Paraashar\footnote{Indian Institute of Technology Hyderabad. Email:{\tt manaswi@cse.iith.ac.in}}
 \and
 Nitin Saurabh\footnote{Indian Institute of Technology Hyderabad. Email:{\tt nitin@cse.iith.ac.in}}
}

\date{}

\begin{document}
\pagenumbering{arabic}
\pagestyle{plain}

\maketitle
\thispagestyle{plain}

\begin{abstract}
\normalsize
    In the world of quantum computing, quantum Fourier transform (QFT) has been one of the most important subroutines for quantum algorithms, as exemplified by Shor's factoring algorithms and phase estimation (which in itself has been used as a primitive in various quantum algorithms). In fact most of the well known quantum algorithms can be divided into two classes, QFT based algorithms and search based algorithms.

The usual circuit for QFT (covered in almost every basic quantum algorithms course) needs $O(q^2)$ gates (and $O(q)$ qubits) to perform the exact Fourier transform over $\mathbb{F}_{2^q}$. Cleve and Watrous (FOCS 2000) improved this gate count to $O(q (\log q)^2 (\log \log q))$ using the integer multiplication algorithm in $O(n (\log n) (\log \log n))$ time for two $n$-bit integers (best known at that time). Our first main result shows that this gate count can be improved to $O(q (\log q))$. In fact, if there is a multiplication circuit, to multiply two $n$-bit integers, with $O(n (\log n)^{\alpha})$ gates; we get a QFT circuit with $O(q (\log q)^{\alpha})$ gates (improving upon $O(q (\log q)^{\alpha + 1})$) gates from the construction of Cleve and Watrous). Given the recent claim by OpenAI about integer multiplication, this opens up the possibility of having a QFT circuit with $o \left( q (\log q)\right)$ gates. 

On approximating the QFT over $\ZZ/(2^q \ZZ)$, the result of Ahokas, Cleve and Hales (ERATO 2003) showed that an approximating circuit can be built with $O \left( q (\log \log(q/\varepsilon))^{\alpha+1} \right)$ gates, where $\epsilon$ is the approximating parameter. Our second main result again improves this to $O \left( q (\log \log(q/\varepsilon))^{\alpha} \right)$ gates.

In comparison to the previous results mentioned above (Cleve and Watrous; Ahokas, Cleve and Hales), our constructions use an additional single qubit rotation gate whose angle of rotation depends on $q$. Due to the presence of this gate, we are not able to show that our circuit families are uniform.

\end{abstract}




\section{Introduction}

The quantum Fourier transform (QFT) is a fundamental quantum algorithm and
it has applications all over quantum computing. In particular, it plays a central role in Shor's algorithms for factoring and discrete
logarithms~\cite{Shor1997}, and in the standard quantum phase-estimation
procedure~\cite{CEMM1998}.
We refer the reader to~\cite{NielsenChuang, deWolf, ChildsVanDam2010, CEMM1998} for an introduction to the QFT and its applications.

For an integer $0\le x<2^q$, let $\ket{x}$ denote its $q$-qubit computational basis encoding.
The QFT is the unitary $F_q$ defined by
\begin{equation*}
F_q \ket{x}=2^{-q/2} \sum_{y=0}^{2^q-1}e^{2\pi ixy/2^q}\ket{y},
\qquad \text{ for all } 0 \leq x<2^q.
\end{equation*}
By linearity, the action of $F_q$ extends to any $q$-qubit state $\ket{\psi}$.

Let $U$ be a unitary circuit acting on the $q$-qubit input register and $a$ auxiliary qubits initialized to $\ket{0^a}$.
We say that $U$ implements $F_q$ exactly if
\begin{equation*}
U\bigl(\ket{\psi}\ket{0^a}\bigr)
=(F_q\ket{\psi})\ket{0^a}
\end{equation*}
for every $q$-qubit state $\ket{\psi}$.
We say that $U$ is an $\varepsilon$-approximate implementation of $F_q$ if
\begin{equation}\label{eq:intro-error}
\sup_{\norm{\psi}=1}
\norm{U\bigl(\ket{\psi}\ket{0^a}\bigr)
      -(F_q\ket{\psi})\ket{0^a}}
\le\varepsilon.
\end{equation}
The norm above is the Euclidean norm.\\

In this paper, we study the complexity of computing QFT in the quantum circuit model. Specifically, we are interested in the following question: \emph{how many gates are needed to implement the transform, both exactly as well as approximately?}

\paragraph{Why look at approximate QFT?} Approximation is a natural part of implementing the QFT.
Even in an exact circuit, the required elementary gates may themselves needs to be approximated.

For example, fault-tolerant implementations over Clifford+$T$ must account
for the cost of synthesizing these gates~\cite{NamSuMaslov2020}.
Moreover, QFT is almost exclusively used as a subroutine. In that case an algorithm using the QFT generally requires only enough accuracy from QFT to preserve its overall success probability.
Physical noise gives a further reason to study this tradeoff:
a longer circuit that more closely approximates the ideal transform
need not produce a more accurate final computation.
This was demonstrated by Barenco, Ekert, Suominen, and T{\"o}rm{\"a}~\cite{BEST96} (building on Coppersmith's truncated QFT construction~\cite{Cop94}) for periodicity
estimation under their model of decoherence.

\subsection{Previous work}\label{sec:previous}

There has been a considerable amount of work in making quantum Fourier transform efficient in terms of the number of gates used.
The standard circuit of the quantum Fourier transform (QFT), done by Coppersmith~\cite{Cop94}, is comprised of $O(q)$ Hadamard
gates and $O\left(q^2 \right)$ controlled phase gates, giving a total of $O(q^2)$ gates.

In the same paper, Coppersmith~\cite{Cop94} introduced the notion of approximate QFT, and obtained the approximate QFT circuit by omitting sufficiently small rotations. This reduces the gate count to $O(q\log(q/\epsilon))$, where $\epsilon$ denotes
the approximation error. 

To reduce the gate count in exact QFT, Cleve and Watrous~\cite{CleveWatrous2000} developed a different approach by replacing groups of controlled rotations with reversible integer multiplication. Their construction yields an exact QFT on $q$ qubits using
\(
  O\!\left(q(\log q)^{2} \log \log q \right)
\)
elementary quantum gates and Approximate QFT using $O(q\log(q/\epsilon))$ many gates.
This improves to 
\(
  O\!\left(q(\log q)^{2} \right)
\)
assuming the recent integer multiplication algorithm by Harvey and
van der Hoeven~\cite{HarveyVanDerHoeven2021}. Whether exact QFT on $q$-qubits can be implemented by a circuit of size $O(q)$ has been posed as an open problem by Aaronson~\cite{Aaronson2025QFT}.

On the approximate QFT front, Ahokas, Cleve, and Hales~\cite{AhokasCleveHales} gave a construction using \(
  O\!\left(q(\log\log (q/\epsilon))^2\log\log\log (q/\epsilon)\right) \)
elementary gates for approximation error $\epsilon$. A detailed exposition could be found in Ahokas's thesis~\cite[Section~3.1]{Aho04}, which also provides an important comparison with \cite{CleveWatrous2000}: integer multiplication can be used to improve both exact and approximate QFT constructions, with approximation permitting a smaller asymptotic gate count. 

The result of Cleve and Watrous~\cite{CleveWatrous2000} discussed above also shows that the QFT can be approximated in depth $O(\log q)$ for inverse-polynomial error. 
For the same error, when unbounded quantum fan-out gates are available,
H{\o}yer and \v{S}palek~\cite{HoyerSpalek2005} showed that the QFT can be approximated by constant-depth, polynomial-size circuits. Mosca and Zalka~\cite{MoscaZalka2003} used amplitude amplification to obtain an exact QFT over $\mathbb Z/N\mathbb Z$ for arbitrary $N$.

There has been work on reducing the number of ancilla qubits too, recently Kahanamoku-Meyer and Yao~\cite{KahanamokuMeyerYao2024} gave an exact QFT with \(O(q^{1+\delta})\) gates for every fixed \(\delta>0\), using \emph{zero ancillas}. Here \(\delta\) controls the complexity exponent and is distinct from the approximation error. Compared with the polylogarithmic-overhead bound of \cite{CleveWatrous2000}, this result emphasizes reduced workspace: the product need not be explicitly stored in an ancillary register before phases are applied and the product is uncomputed.

\subsection{Our results}\label{sec:results}
Cleve and Watrous~\cite{CleveWatrous2000} showed how to use a classical circuit for
integer multiplication to construct an exact QFT circuit.
If two $n$-bit integers can be multiplied by Boolean circuits over two-input AND and OR gates
and single-input NOT gates, with unrestricted fan-out, using
\begin{equation}\label{eq:multiplication-cost}
M(n) = O \left( n (\log n)^{\alpha} \right)
\end{equation}
gates for a fixed constant $\alpha>0$, the construction of~\cite{CleveWatrous2000}  gives an exact QFT on $q$ qubits
using $O \left( q (\log q)^{\alpha+1} \right)$ gates.
For approximation error $0<\varepsilon<1$, the approach of Ahokas, Cleve, and
Hales~\cite[Sections~3.1.1--3.1.2]{Aho04} retains only $O \left( \log(q/\varepsilon) \right)$
leading bits of each input to a multiplication.
Assuming the multiplication bound in Equation~\ref{eq:multiplication-cost},
this gives an approximate QFT using
$O \left( q (\log \log(q/\varepsilon))^{\alpha+1} \right)$ gates.
Our main results improve these gate bounds to
\begin{equation*}
O \left( q (\log q)^{\alpha} \right)
\quad \text{and} \quad
O \left( q (\log \log(q/\varepsilon))^{\alpha} \right)
\end{equation*}
for exact and $\varepsilon$-approximate circuits, respectively.
For the exact QFT, we use the same gate set as
Cleve and Watrous~\cite{CleveWatrous2000}, together with an additional
single-qubit rotation gate whose angle depends on $q$. This non-standard gate is described below. 
For the $\varepsilon$-approximate QFT, our construction uses the same gate set as in
Cleve and Watrous.

Both theorems hold for every integer $q\geq64$.
The exact construction does not require $q$ to be a power of two.
The approximate construction allows unequal recursive splits.
All logarithms have base two.

\begin{theorem}[Exact QFT]\label{thm:exact}
Assume the multiplication bound in Equation~\ref{eq:multiplication-cost}.
For every integer $q \geq 64$, the transform $F_q$ has an exact implementation using
\begin{align*}
    O \left( q (\log q)^{\alpha} \right)
\quad \text{gates and auxiliary qubits}.
\end{align*}
All auxiliary qubits return to zero.
\end{theorem}

The exact implementation uses Hadamard gates, controlled-NOT gates, and phase gates
\begin{equation}\label{eq:phase-gates}
P \left( \frac{2\pi a}{2^d} \right)
= \begin{pmatrix}
1 & 0 \\
0 & e^{2\pi i a / 2^d}
\end{pmatrix},
\qquad a \in \mathbb{Z}, \quad d \in \mathbb{Z}_{\geq 1}.
\end{equation}
In addition, it uses a single-qubit rotation 
\begin{equation}\label{eq:additional-rotation}
B \left( \theta_q \right)
= \begin{pmatrix}
\cos \theta_q & -\sin \theta_q \\
\sin \theta_q & \cos \theta_q
\end{pmatrix},
\end{equation}
whose angle depends only on $q$.

\begin{theorem}[Approximate QFT]\label{thm:approx}
Assume the multiplication bound in Equation~\ref{eq:multiplication-cost}.
For every integer $q \geq 64$ and every $0<\varepsilon<1$, let
\begin{equation*}
L = \log(q/\varepsilon).
\end{equation*}
The transform $F_q$ has an $\varepsilon$-approximate implementation using
\begin{equation*}
O \left( q (\log L)^{\alpha} \right)
\quad \text{gates and} \quad
O \left( L (\log L)^{\alpha} \right)
\quad \text{auxiliary qubits}.
\end{equation*}
The implicit constants may depend on $\alpha$ and the multiplication family,
but not on $q$ or $\varepsilon$.
\end{theorem}

The implementation in Theorem~\ref{thm:approx} uses Hadamard, controlled-NOT,
and exact phase gates $P \left( 2\pi a / 2^d \right)$ for all $a \in \mathbb{Z}$ (the $a$ used can depend on $\theta_q$). Presently we are not able to compute the angle $\theta_q$ efficiently using a classical algorithm, thus our construction of the circuit families in both the theorems above are non-uniform. This is a natural open question from this work.

Our results thus improve the connection between the costs of integer multiplication and the QFT.
In~\cite{OpenAI2026Multiplication}, OpenAI proposed an integer multiplication algorithm
with complexity $O\left(n(\log n)^{1-\kappa}\right)$ for a fixed constant $0<\kappa<1$.
Harvey~\cite{Harvey2026Multiplication} gave an exposition of the proposed algorithm
and discusses this bound in the Boolean circuit model.
Assuming the claimed bound holds for the Boolean circuits specified in
Equation~\ref{eq:multiplication-cost}, we may take $\alpha=1-\kappa$ in our theorems.
Theorem~\ref{thm:exact} then gives an exact QFT using
$O\left(q(\log q)^{1-\kappa}\right)=o(q\log q)$ gates,
with exact access to the additional rotation $B(\theta_q)$ specified above.
Theorem~\ref{thm:approx} similarly gives an approximate QFT using
$O\left(q(\log\log q)^{1-\kappa}\right)=o(q\log\log q)$ gates
for inverse-polynomial error, without requiring that additional exact rotation. Whether the exact or approximate QFT on $q$ qubits admits a circuit of size $O(q)$
remains an open question (see~\cite{Aaronson2025QFT}).

\subsection{Proof ideas}\label{sec:proof-ideas-short}

To prove Theorem~\ref{thm:exact}, we combine the exactification method of
Mosca and Zalka~\cite{MoscaZalka2003} with the multiplication-based implementation
of the QFT due to Cleve and Watrous~\cite{CleveWatrous2000}.
We then use the resulting exact circuit in the approximate construction of
Ahokas, Cleve, and Hales~\cite{AhokasCleveHales},~\cite[Theorem~3.1]{Aho04} to prove Theorem~\ref{thm:approx}. The construction is due to Ahokas, Cleve, and Hales~\cite{AhokasCleveHales}. Throughout, we refer to their work for the construction and cite the relevant sections of Ahokas's thesis~\cite{Aho04} for the formal statements of the results we use.

\paragraph{Exact QFT.}
Mosca and Zalka~\cite[Section~2]{MoscaZalka2003} turn an approximate QFT circuit to an exact one
by first making its success probability independent of the input Fourier state.
An additional single-qubit rotation then reduces the success probability to $1/4$,
so that one step of amplitude amplification produces the correct answer with certainty.

In our construction, we first copy the Fourier state using the circuit of Cleve and
Watrous, then apply the approximate inverse QFT to the copy~\cite[Section~3.2]{CleveWatrous2000}.
This leaves the original Fourier state available for the checking during amplitude amplification.

Besides calls to the approximate circuit and its inverse, the construction
requires operations on Fourier states.
These are carried out using a constant number of
the product-phase operations defined below, together with $O(q)$ additional gates.
For $q$-bit integers $a$ and $b$, the product-phase operation is
\begin{equation}\label{eq:product-phase-short}
\ket{a}\ket{b}\longmapsto e^{\pm2\pi iab/2^q}\ket{a}\ket{b}.
\end{equation}
This operation can be implemented using the multiplication-based construction
of Cleve and Watrous~\cite[Section~4]{CleveWatrous2000},
at a cost of $O(M(q))$ gates.
Thus, if the approximate circuit costs $G(q)$ gates, the exact circuit costs $O(G(q)+M(q))$ gates.

We choose the approximate circuit of Ahokas, Cleve, and Hales~\cite[Theorem~3.1]{Aho04}.
Under the multiplication assumption in Equation~\ref{eq:multiplication-cost}, it uses
$G(q)=O\left(q(\log\log q)^{\alpha+1}\right)$ gates at error at most $1/2$. The resulting exact QFT uses $O\left(q(\log q)^\alpha\right)$ gates.

Lemma~\ref{lem:exact-from-recovery} and it's proof idea explains the construction in detail,
including the implementation of these operations and their gate costs.

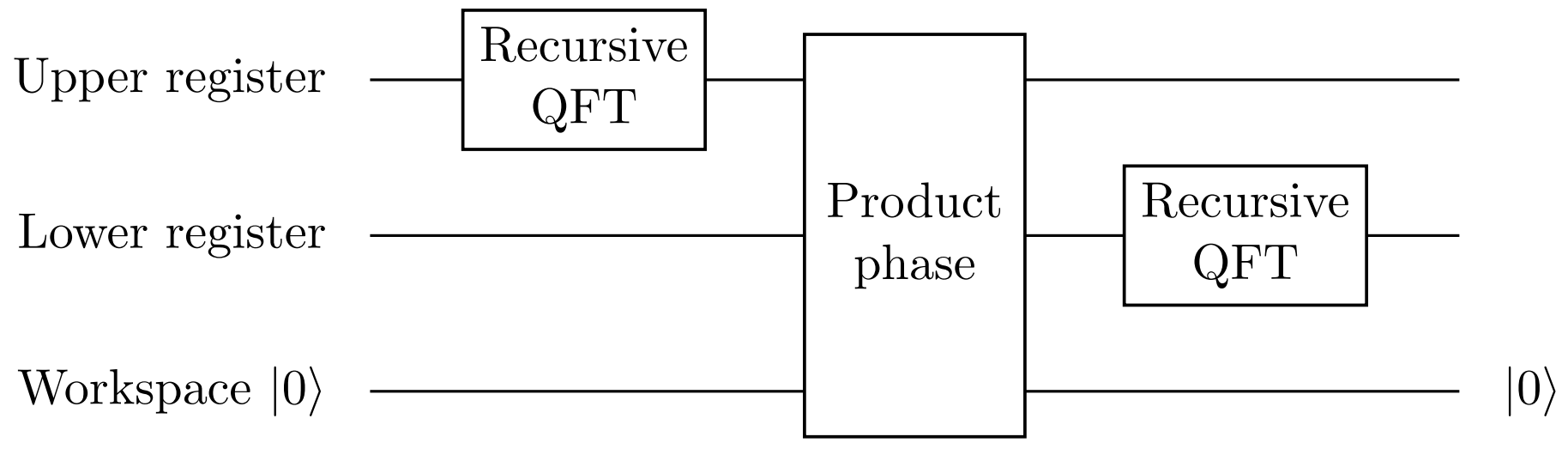
\begin{figure}[htbp]
\centering
\input{figures/ach-recursion.tex}
\caption{One recursive step of the ACH construction.}
\label{fig:ach-recursion-short}
\end{figure}

\paragraph{Using the improved exact circuit for approximation.}
We use the recursive construction of Ahokas, Cleve, and
Hales~\cite[Section~3.1.1]{Aho04} to prove Theorem~\ref{thm:approx}.
Figure~\ref{fig:ach-recursion-short} shows one recursive step of this construction.
Starting with the $q$-qubit input, their construction proceeds as follows.
Split the input into a lower register of $q/2$ qubits
and an upper register of $q/2$ qubits.
Apply the construction recursively to the upper register, apply the product-phase
operation in Equation~\ref{eq:product-phase-short} between the two registers,
and then apply the construction recursively to the lower register.
Finally, exchange the two registers to obtain the required output order. For the product-phase operation, multiplication is done only on the highest $k$ bits of each
multiplication input, where $k=\Theta(\log(q/\varepsilon))$.
Stop the recursion when a register has at most $2k$ qubits, and apply an exact QFT
to that register.

Thus there are $O(q/k)$ base cases and $O(q/k)$ product-phase operations.
Each product-phase operation costs $O\left(k(\log k)^\alpha\right)$ gates.
Using the exact construction of Cleve and Watrous~\cite{CleveWatrous2000},
each base case costs $O\left(k(\log k)^{\alpha+1}\right)$ gates.
The base cases therefore determine the overall bound of
$O\left(q(\log k)^{\alpha+1}\right)$ gates.

We replace these base cases by our improved exact QFT, which costs
$O\left(k(\log k)^\alpha\right)$ gates per call and
the total cost becomes $O\left(q(\log k)^\alpha\right)$ gates.
Recall that our improved exact circuit uses an additional single-qubit rotation.
We efficiently approximate these rotations using the phase gates $P(2\pi a/2^d)$
defined in Equation~\ref{eq:phase-gates}.
Since $k=\Theta(\log(q/\varepsilon))$, the resulting circuit uses
$O\left(q(\log\log(q/\varepsilon))^\alpha\right)$ gates.

Section~\ref{sec:proof-approx} gives the full construction and proves
its approximation error, gate count, and auxiliary-space bounds.

\section{Preliminaries}\label{sec:preliminaries}
Throughout this paper, we follow the standard notation and conventions of quantum computation. We refer the reader to Nielsen and Chuang~\cite{NielsenChuang} for background and further details. 

Let $N=2^q$, $\omega=e^{2\pi i/N}$, and $\ket{\phi_x}=F_q\ket{x}$.
Recall that, for $0\leq a<N$, $\ket{a}$ denotes the computational-basis state
whose $q$ bits give the binary representation of $a$.
Addition and subtraction of register values are performed modulo $N$ using $O(q)$ gates~\cite[Section~4.1]{CuccaroDraperKutinMoulton2004}.

We use the gate model stated after Theorem~\ref{thm:exact}.
NOT and Toffoli gates have constant-size implementations in this model~\cite{NielsenChuang}.
A phase change conditioned on values of $q$ qubits uses $O(q)$ gates
and auxiliary qubits: compute the conjunction of the required bit values,
apply the phase, and uncompute the conjunction.
The same construction computes and uncomputes the OR of $q$ bits with $O(q)$ gates
and auxiliary qubits, all returned to zero.

We also use the standard reversible simulation of a Boolean circuit.
A circuit with $G$ gates, where $G$ is at least the number of input and output bits,
can be implemented reversibly using $O(G)$ gates and wires.
The simulation retains the inputs and clears its temporary registers by copying
the output and reversing the computation~\cite{Bennett1973, NielsenChuang}.

The following lemma implements a product-phase operation using integer multiplication.

\begin{lemma}[Product-phase operation {\cite{Cleve1994}\cite[Section~4]{CleveWatrous2000}}]\label{lem:product-phase}
Let $A$ and $B$ be integers stored in two $q$-qubit registers, and let $1\leq d\leq2q$.
The operation
\begin{equation*}
\ket{A}\ket{B}\longmapsto
 e^{\pm2\pi iAB/2^d}\ket{A}\ket{B}
\end{equation*}
has an exact implementation using $O(M(q))$ gates and auxiliary qubits.
All auxiliary qubits return to zero.
\end{lemma}

\begin{proof}
Compute the product $AB$ into a $2q$-qubit auxiliary register initialized to zero,
using the reversible circuit for multiplication.
Let $AB=\sum_{j=0}^{2q-1}p_j2^j$.
Apply $P\left(\pm2\pi2^j/2^d\right)$ to each bit $p_j$ with $0\leq j<d$,
and reverse the multiplication.
The phase gates multiply the state by
\begin{equation*}
\prod_{j=0}^{d-1} e^{\pm2\pi i p_j2^j/2^d}
=
e^{\pm2\pi i\left(\sum_{j=0}^{d-1}p_j2^j\right)/2^d}
=
e^{\pm2\pi iAB/2^d}.
\end{equation*}
The last equality holds because the terms omitted in $AB/2^d$ sum to an integer
and thus do not contribute to the phase.
Since the phase gates leave the product unchanged, reversing the multiplication
clears every temporary register while preserving the desired phase.
The forward and reverse computations cost $O(M(q))$ gates.
\end{proof}

We will use the following result of \cite{CleveWatrous2000}, which shows that Fourier basis states can be copied efficiently. 
\begin{lemma}[Fourier-state copying {\cite[Section~3.2]{CleveWatrous2000}}]\label{lem:fourier-copy}
There is a circuit using $O(q)$ gates and auxiliary qubits that maps
\begin{equation*}
\ket{\phi_x}_I\ket{0^q}_O\longmapsto\ket{\phi_x}_I\ket{\phi_x}_O
\end{equation*}
for every $0\leq x<N$.
All auxiliary qubits return to zero.
\end{lemma}

\begin{proof}
Apply Hadamard gates to $O$, then subtract its value from $I$.
For a fixed value $c$ in $O$, changing variables in the sum gives
\begin{equation}\label{eq:fourier-copy-shift}
\frac1{\sqrt N}\sum_{s=0}^{N-1}\omega^{xs}\ket{s-c}_I\ket{c}_O
=\omega^{xc}\ket{\phi_x}_I\ket{c}_O.
\end{equation}
Thus the uniform superposition in $O$ acquires the phases of $\ket{\phi_x}$.
The Hadamard gates and reversible subtraction use $O(q)$ gates and auxiliary qubits,
with temporary qubits returned to zero.
\end{proof}

We will also use the following result from \cite{MoscaZalka2003}, which allows us to apply a conditional phase based on whether a Fourier-state label matches a computational-basis label.

\begin{lemma}[Fourier-state answer test {\cite[Section~2.1.1]{MoscaZalka2003}}]\label{lem:fourier-answer-test}
Let $I$ and $O$ be $q$-qubit registers and $E$ a single qubit.
There is a unitary $D$ satisfying
\begin{equation*}
D\left(\ket{\phi_x}_I\ket{y}_O\ket{b}_E\right)
=(-1)^{\mathbf{1}_{\{x=y,\ b=1\}}}
\ket{\phi_x}_I\ket{y}_O\ket{b}_E
\end{equation*}
for all $0\leq x,y<N$ and $b\in\{0,1\}$.
Here the exponent is one when both $x=y$ and $b=1$, and zero otherwise.
The operation uses $O(M(q))$ gates and auxiliary qubits. All auxiliary qubits return to zero.
\end{lemma}

\begin{proof}
The idea is to transform $I$ so that equality $x=y$ can be tested by computing
the OR of its $q$ bits, apply the required phase when this OR is zero,
and then undo the computation.

First apply the product phase $\omega^{-ys}$ using Lemma~\ref{lem:product-phase}
with $d=q$, where $y$ is the value in $O$ and $s$ is the computational-basis value in $I$.

For fixed $x,y,b$, the action on the three registers is

\begin{equation*}
\begin{aligned}
\ket{\phi_x}_I\ket{y}_O\ket{b}_E
&=\frac1{\sqrt N}\sum_{s=0}^{N-1}\omega^{xs}\ket{s}_I\ket{y}_O\ket{b}_E\\
&\longmapsto\frac1{\sqrt N}\sum_{s=0}^{N-1}
\omega^{xs}\omega^{-ys}\ket{s}_I\ket{y}_O\ket{b}_E\\
&=\frac1{\sqrt N}\sum_{s=0}^{N-1}
\omega^{(x-y)s}\ket{s}_I\ket{y}_O\ket{b}_E\\
&=\ket{\phi_{x-y}}_I\ket{y}_O\ket{b}_E.
\end{aligned}
\end{equation*}
Thus the product phase subtracts the value $y$ from the Fourier label $x$.

Next apply $H^{\otimes q}$ to $I$.
If $x=y$, then $\ket{\phi_{x-y}}=\ket{\phi_0}$ is the uniform superposition,
so these Hadamard gates map it to $\ket{0^q}$.
If $x\ne y$, the resulting state is orthogonal to $\ket{0^q}$, since
\begin{equation*}
\bra{0^q}H^{\otimes q}\ket{\phi_{x-y}}
=\frac1N\sum_{s=0}^{N-1}\omega^{(x-y)s}=0.
\end{equation*}

Compute the OR of the bits in $I$ into an auxiliary qubit and apply a phase of $-1$
when this qubit is zero and $E$ is one.
This gives the required phase precisely when $x=y$ and $b=1$.

Uncompute the OR, the Hadamard gates, and the product phase to restore the input
registers and clear all auxiliary qubits while retaining the required phase.

The product phase and its inverse use $O(M(q))$ gates and auxiliary qubits
by Lemma~\ref{lem:product-phase}.
The Hadamard gates and conditional phase change add $O(q)$ gates and auxiliary qubits,
so the total cost is $O(M(q))$.
\end{proof}

The next lemma gives the action of one Grover iterate when the initial probability
of the good states is $1/4$: this choice makes the final probability exactly one.

\begin{lemma}[One Grover iterate {\cite{MoscaZalka2003,BrassardHoyerMoscaTapp2002}}]\label{lem:one-step-amplification}
Let $U$ be a unitary on $m$ qubits such that
\begin{equation*}
U\ket{0}=\ket{s_x}=\frac12\ket{g_x}+\frac{\sqrt3}{2}\ket{h_x},
\end{equation*}
where $\ket{0}$ denotes the all-zero state and $\ket{g_x}$ and $\ket{h_x}$
are orthogonal unit vectors.
Let $D$ be a unitary satisfying
\begin{equation*}
D\ket{g_x}=-\ket{g_x},\qquad D\ket{h_x}=\ket{h_x}.
\end{equation*}
Write $R_0=2\ket{0}\bra{0}-\mathbb{I}$.
Then
\begin{equation*}
UR_0U^\dagger D\ket{s_x}=\ket{g_x}.
\end{equation*}
This Grover iterate uses one call each to $D,U^\dagger,R_0,U$.
The reflection $R_0$ uses $O(m)$ gates and $O(m)$ auxiliary qubits,
all returned to zero.
\end{lemma}

\begin{proof}
Since $U\ket{0}=\ket{s_x}$,
$UR_0U^\dagger=2\ket{s_x}\bra{s_x}-\mathbb{I}$.
The action of $D$ gives
\begin{equation*}
D\ket{s_x}=-\frac12\ket{g_x}+\frac{\sqrt3}{2}\ket{h_x}.
\end{equation*}
Since $\ket{g_x}$ and $\ket{h_x}$ are orthogonal unit vectors,
$\bra{s_x}D\ket{s_x}=-1/4+3/4=1/2$.
Therefore
\begin{equation*}
\begin{aligned}
(2\ket{s_x}\bra{s_x}-\mathbb{I})D\ket{s_x}
&=\ket{s_x}-D\ket{s_x}\\
&=\ket{g_x}.
\end{aligned}
\end{equation*}

To implement $R_0$, compute the OR of the $m$ bits into an auxiliary qubit,
apply a $Z$ gate to that qubit, and uncompute the OR.
This leaves $\ket{0}$ unchanged
and multiplies every other computational-basis state by $-1$.
\end{proof}

\begin{theorem}[Approximate QFT, {\cite[Theorem~3.1]{Aho04}}]\label{thm:ach-approx}
Assume the multiplication bound in Equation~\ref{eq:multiplication-cost}.
For every integer $q\geq64$, there is a unitary circuit $W$ on $q$ qubits such that
\begin{equation*}
\norm{W-F_q}\leq\frac12,
\end{equation*}
where the norm is the operator norm.
The circuit uses $O\left(q(\log\log q)^{\alpha+1}\right)$ gates, consisting of the phase gates
in Equation~\ref{eq:phase-gates} and Hadamard and controlled-NOT gates,
and $O\left(q(\log\log q)^{\alpha+1}\right)$ auxiliary qubits, all returned to zero.
The inverse circuit has the same resource bounds and approximates $F_q^\dagger$
to the same error.
\end{theorem}

\begin{proof}
We apply the multiplication bound in
Equation~\ref{eq:multiplication-cost} to the ACH construction~\cite[Sections~3.1.1--3.1.3]{Aho04}, described in
Section~\ref{sec:proof-ideas-short}.

Let $k=\Theta(\log q)$ as in Section~\ref{sec:proof-ideas-short}. Their construction uses $O(q/k)$ product-phase operations
and exact QFTs on at most $2k$ qubits.
By Lemma~\ref{lem:product-phase} and the Cleve--Watrous exact
construction~\cite[Section~4]{CleveWatrous2000}, these cost
$O\left(k(\log k)^\alpha\right)$ and
$O\left(k(\log k)^{\alpha+1}\right)$ gates, respectively.
Their total cost is therefore $O\left(q(\log\log q)^{\alpha+1}\right)$.
The error analysis in~\cite[Section~3.1.2, Equation~(3.5)]{Aho04}
shows that choosing a suitable $k=\Theta(\log q)$ gives approximation error at most $1/2$.
Furthermore, the circuits return their auxiliary qubits to zero, and the number
of auxiliary qubits is bounded by a constant times the gate count.
Reversing the circuit preserves these bounds and the operator-norm error.
\end{proof}

\section{Proof of Theorem~\ref{thm:exact}}\label{sec:proof-exact}

Let $V$ be a unitary circuit on $q$ qubits satisfying
$\norm{V-F_q^\dagger}\leq1/2$ in operator norm.
Suppose $V$ can be implemented using $G(q)$ gates and $a(q)$ auxiliary qubits that are
returned to zero.
For $0\leq x<N$, define
\begin{equation}\label{eq:recovery-average}
p_x=\left|\bra{x}V\ket{\phi_x}\right|^2,
\qquad
p=\frac1N\sum_{x=0}^{N-1}p_x.
\end{equation}
Thus $p_x$ is the probability of obtaining $x$ when $V$ is applied to
$\ket{\phi_x}$ and the output is measured in the computational basis.
The quantity $p$ is the average of these probabilities over uniformly chosen $x$.
Since $\norm{V\ket{\phi_x}-\ket{x}}\leq1/2$, the component orthogonal to
$\ket{x}$ has squared norm at most $1/4$.
Thus $p_x\geq3/4$ for every $x$, and $p\geq3/4$.

We first modify the circuit so that the probability of the correct answer equals
$p$ for every $x$.
To make this answer certain with one Grover iterate, we introduce an extra qubit
and require both a correct answer and value one in that qubit.
The joint probability must be $1/4$, so we choose the probability of value one to be $1/(4p)$.
Since $p\geq3/4$, this choice is possible. Let
\begin{equation}\label{eq:exact-rotation-angle}
\theta_q=\arcsin\left(\frac1{2\sqrt p}\right).
\end{equation}
We assume exact access to the rotation $B\left(\theta_q\right)$ defined in
Equation~\ref{eq:additional-rotation}, at unit cost.

The angle depends on $V$ and $q$, and is independent of the input $x$.
Its inverse has constant cost as well, since
$B\left(\theta_q\right)^\dagger=X B\left(\theta_q\right)X$.

\begin{lemma}\label{lem:exact-from-recovery}
Under these assumptions, $F_q$ and $F_q^\dagger$ have exact implementations using
\begin{equation*}
O\left(G(q)+M(q)\right)\quad\text{gates}
\qquad\text{and}\qquad
O\left(a(q)+M(q)\right)\quad\text{auxiliary qubits}.
\end{equation*}
All auxiliary qubits return to zero.
\end{lemma}

\begin{proof}[Proof of Lemma~\ref{lem:exact-from-recovery}]

Let $I$ be the input register, initially containing $\ket{\phi_x}$.
All auxiliary registers introduced below are initialized to zero.

\paragraph{Preparing a state with success probability $1/4$.}\label{proc:prepare-answer}
We first modify the approximate circuit so that it produces the correct answer
with the same probability for every $x$.
Following Mosca and Zalka~\cite[Section~2.2]{MoscaZalka2003}, we do this by
averaging over a uniformly chosen shift of the Fourier label.

Introduce a $q$-qubit output register $O$.
Apply Lemma~\ref{lem:fourier-copy} to copy the Fourier state into $O$,
obtaining $\ket{\phi_x}_I\ket{\phi_x}_O$ at a cost of $O(q)$ gates.
We retain the original Fourier state in $I$ to check the answer later. Add another $q$-qubit register $S$ and apply the following operations:
\begin{enumerate}[label=\arabic*.]
\item Apply Hadamard gates to $S$, preparing a uniform superposition of shifts $r$.
\item Apply the product phase $\omega^{-rs}$, where $r$ and $s$ are the values
in $S$ and $O$, respectively.
Expanding the Fourier state in $O$ shows how this shifts its label:
\begin{equation*}
\begin{aligned}
\frac1{\sqrt N}\sum_{r=0}^{N-1}
\ket{\phi_x}_I\ket{\phi_x}_O\ket{r}_S
&=\frac1N\sum_{r,s=0}^{N-1}
\omega^{xs}\ket{\phi_x}_I\ket{s}_O\ket{r}_S\\
&\longmapsto\frac1N\sum_{r,s=0}^{N-1}
\omega^{xs}\omega^{-rs}\ket{\phi_x}_I\ket{s}_O\ket{r}_S\\
&=\frac1N\sum_{r,s=0}^{N-1}
\omega^{(x-r)s}\ket{\phi_x}_I\ket{s}_O\ket{r}_S\\
&=\frac1{\sqrt N}\sum_{r=0}^{N-1}
\ket{\phi_x}_I\ket{\phi_{x-r}}_O\ket{r}_S.
\end{aligned}
\end{equation*}
Thus, for each shift $r$, the state in $O$ becomes $\ket{\phi_{x-r}}$,
while the original Fourier state in $I$ is unchanged.
\item Apply $V$ to $O$, then add the value in $S$ to $O$ modulo $N$.
\end{enumerate}
For each shift $r$, the answer in $O$ equals $x$ precisely when $V$ gives
$x-r$ on input $\ket{\phi_{x-r}}$, which has probability $p_{x-r}$.
Thus the probability that $O$ contains $x$ is
\begin{equation*}
\frac1N\sum_{r=0}^{N-1}p_{x-r}=p.
\end{equation*}

Finally introduce an auxiliary qubit $E$ to adjust the success probability to $1/4$.
Apply $B(\theta_q)$ to $E$, obtaining
\begin{equation*}
B(\theta_q)\ket{0}_E
=\sqrt{1-\frac1{4p}}\ket{0}_E+\frac1{2\sqrt p}\ket{1}_E.
\end{equation*}
For fixed $x$, call a state $\ket{\phi_x}_I\ket{y}_O\ket{r}_S\ket{b}_E$
good when $y=x$ and $b=1$.
The total probability of the good states is $p/(4p)=1/4$,
as in Mosca and Zalka~\cite[Section~4.2]{MoscaZalka2003}.

Let $U$ be the circuit consisting of the operations above and denote the state that it prepares by
\begin{equation*}
\ket{s_x}=U\left(\ket{\phi_x}_I\ket{0}_{O,S,E}\right).
\end{equation*}
It uses one call to $V$, one product-phase operation, and $O(q)$ other gates,
so its cost is $O(G(q)+M(q))$.

\paragraph{Making the answer certain.}\label{proc:good-sign}
Let $D$ be the operation of Lemma~\ref{lem:fourier-answer-test} on $I,O,E$,
acting trivially on $S$.
It multiplies the good-state amplitudes by $-1$ using $O(M(q))$ gates.
Let $R_0$ reflect about the all-zero state of $O,S,E$, acting trivially on $I$;
this costs $O(q)$ gates by Lemma~\ref{lem:one-step-amplification}.

For an input Fourier state $\ket{\phi_x}$, the Grover iteration acts within the
subspace in which register $I$ contains $\ket{\phi_x}$.
The prepared state $\ket{s_x}$ lies in this subspace, and each of the operations
$D,U^\dagger,R_0,U$ preserves it.
Indeed, the copying circuit of Lemma~\ref{lem:fourier-copy} and its inverse
preserve $\ket{\phi_x}$ in $I$, while the remaining operations in $U$ act only
on the other registers.
The answer test $D$ restores $\ket{\phi_x}$ after applying the conditional phase,
as shown in Lemma~\ref{lem:fourier-answer-test}, and $R_0$ does not act on $I$.
Within this subspace, the operation $UR_0U^\dagger$
reflects about $\ket{s_x}$.

Since the good-state probability is $1/4$, write
\begin{equation*}
\ket{s_x}=\frac12\ket{g_x}+\frac{\sqrt3}{2}\ket{h_x},
\end{equation*}
where $\ket{g_x}$ and $\ket{h_x}$ are the normalized good and bad components.
Lemma~\ref{lem:one-step-amplification} then gives
\begin{equation}\label{eq:one-grover-iterate}
UR_0U^\dagger D\ket{s_x}=\ket{g_x}.
\end{equation}
Hence $O$ contains $x$ with certainty, while $I$ still contains $\ket{\phi_x}$.

\paragraph{Clearing the computation.}\label{proc:exact-inverse}
Let $T=UR_0U^\dagger D U$ be the circuit that prepares and amplifies the answer.
Since its output register $O$ contains $x$ with certainty, copy $x$ into a new
$q$-qubit register $R$, initialized to zero, and apply $T^\dagger$ to the original
registers.
This restores the state
\begin{equation*}
\ket{\phi_x}_I\ket{x}_R,
\end{equation*}
with all other registers zero.

We now clear $I$ using the value $x$ in $R$.
As in the proof of Lemma~\ref{lem:fourier-answer-test}, applying the product phase
$\omega^{-xs}$, where $s$ is the computational-basis value in $I$, and then Hadamard
gates maps $\ket{\phi_x}_I$ to $\ket{0^q}_I$.
Swapping $I$ and $R$ therefore gives
\begin{equation*}
\ket{\phi_x}_I\ket{0^b}\longmapsto\ket{x}_I\ket{0^b},
\end{equation*}
where $b$ counts all auxiliary qubits.
Since the Fourier states form a basis, linearity shows that this circuit
implements $F_q^\dagger$ exactly.
Reversing it gives $F_q$.

\paragraph{Gate count and auxiliary space.}
Since $T=UR_0U^\dagger D U$ uses a constant number of calls to $U$, $U^\dagger$,
$D$, and $R_0$, it costs $O(G(q)+M(q))$ gates.
Applying $T^\dagger$ and clearing the remaining registers preserves this bound.
The construction uses $O(a(q)+M(q))$ auxiliary qubits, all returned to zero.
\end{proof}

\begin{proof}[Proof of Theorem~\ref{thm:exact}]
Choose $V$ to be the approximate inverse QFT supplied by Theorem~\ref{thm:ach-approx}.
It has error at most $1/2$ and uses
\begin{equation}\label{eq:ach-qft-cost}
G(q)=O\left(q(\log\log q)^{\alpha+1}\right)
\end{equation}
gates and $O(G(q))$ auxiliary qubits.
Applying Lemma~\ref{lem:exact-from-recovery} gives an exact QFT using
\begin{equation*}
O\left(M(q)+q(\log\log q)^{\alpha+1}\right)
=O\left(q(\log q)^\alpha\right)
\end{equation*}
gates and auxiliary qubits. All auxiliary qubits are returned to zero.
\end{proof}

\section{Proof of Theorem~\ref{thm:approx}}\label{sec:proof-approx}

Let $\mathsf{ExactQFT}$ denote the exact QFT circuit constructed in
Section~\ref{sec:proof-exact} and let $L=\log(q/\varepsilon)$.
We prove Theorem~\ref{thm:approx} in two parts. First, we use $\mathsf{ExactQFT}$ in the ACH construction (see Section~\ref{sec:proof-ideas-short}) to obtain an approximate
QFT with $O\left(q(\log L)^\alpha\right)$ gates and
$O\left(L(\log L)^\alpha\right)$ auxiliary qubits.This circuit still uses the additional rotations of the form given in
Equation~\ref{eq:additional-rotation}. In our second step, we approximate these rotations using Hadamard gates and the phase gates
$P(2\pi a/2^d)$.
We show that the resulting circuit has error at most $\varepsilon$ without
changing the gate or auxiliary-qubit bounds.

Fix $q\geq64$ and $0<\varepsilon<1$.
Choose an integer
\begin{equation}\label{eq:hand-approx-cutoff}
k=\Theta(L),
\end{equation}
with a sufficiently large constant.

\subsection{Approximate QFT using exact base cases}\label{sec:hand-approx-circuit}

Consider an $n$-qubit register with $n>2k$.
Split it into a lower register of $r=\lfloor n/2\rfloor$ qubits and an upper
register of $s=\lceil n/2\rceil$ qubits.
The full QFT first transforms the upper register, applies
\begin{equation*}
\ket{u}\ket{v}\longmapsto e^{2\pi iuv/2^n}\ket{u}\ket{v},
\end{equation*}
then transforms the lower register and exchanges the two registers.
Here $u$ is the value in the lower $r$-qubit register and $v$ is the value in the
upper $s$-qubit register at the time of the product-phase operation.
This is the Cleve--Watrous recursion~\cite[Section~4]{CleveWatrous2000}.

ACH keep only the highest $k$ bits of each value
when calculating the product phase.
Since $n>2k$, both registers have at least $k$ qubits.
Write their retained values as $u'$ and $v'$.
In the phase calculation, replace $u$ and $v$ by $2^{r-k}u'$ and $2^{s-k}v'$,
respectively, to obtain the shortened operation below
\begin{equation}\label{eq:hand-short-product}
\ket{u}\ket{v}\longmapsto
 e^{2\pi i u'v'/2^{2k}}\ket{u}\ket{v}.
\end{equation}
Apply Lemma~\ref{lem:product-phase} to the two groups of $k$ selected qubits,
with $q=k$ and $d=2k$.
This implements the operation with $O(M(k))$ gates and auxiliary qubits,
all returned to zero, and leaves the unselected bits untouched.

When $n\leq2k$, we apply $\mathsf{ExactQFT}$ to the entire $n$-qubit register.
The procedure $\mathsf{ApproxQFT}$ below gives the complete construction.

\paragraph{Procedure $\mathsf{ApproxQFT}$.}\label{proc:hand-approx-qft}
Apply the following procedure to an $n$-qubit register, initially with $n=q$,
keeping $k$ fixed throughout.
All calls share an auxiliary register $A$, initialized to $\ket{0^a}$,
and return it to zero.
\begin{enumerate}[label=\arabic*.]
\item If $n\leq2k$, apply $\mathsf{ExactQFT}$ on these $n$ qubits and stop.
\item Otherwise split the register into its lower $r$ qubits and upper $s$ qubits,
as above.
\item Apply $\mathsf{ApproxQFT}$ to the upper register.
\item Apply the shortened product-phase operation in Equation~\ref{eq:hand-short-product}.
\item Apply $\mathsf{ApproxQFT}$ to the lower register.
\item Exchange the registers, placing the transformed lower register before the
transformed upper register.
\end{enumerate}
Let $U_0$ denote the circuit implemented by $\mathsf{ApproxQFT}$.

\paragraph{Correctness.}\label{sec:hand-approx-correctness}
The base-case calls implement exact QFTs, so the only error comes from
shortening the product phases.
To bound one such error, write
\begin{equation*}
u=2^{r-k}u'+a_0,
\qquad v=2^{s-k}v'+b_0,
\qquad 0\leq a_0<2^{r-k},\quad 0\leq b_0<2^{s-k}.
\end{equation*}
The omitted part of the phase exponent is
\begin{equation*}
0\leq\frac{uv}{2^n}-\frac{u'v'}{2^{2k}}
=\frac{a_0v+2^{r-k}u'b_0}{2^n}
<2\cdot2^{-k}.
\end{equation*}
Thus the full and shortened product phases differ by at most $4\pi2^{-k}$
in operator norm, using $\left|e^{ix}-e^{iy}\right|\leq|x-y|$.
There are fewer than $q$ product-phase operations.
Replacing them one at a time and applying the triangle inequality bounds the
error of the shortened transform by $4\pi q2^{-k}$.
Thus, for every input state $\ket{\psi}$,
\begin{equation}\label{eq:hand-shortening-error}
\norm{U_0\left(\ket{\psi}\ket{0^a}\right)
 -(F_q\ket{\psi})\ket{0^a}}
\leq4\pi q2^{-k}.
\end{equation}

\paragraph{Gate count and auxiliary space.}\label{sec:hand-approx-cost}

Suppose first that $q>2k$.
The base-case registers partition the $q$ input qubits, and each contains between
$k$ and $2k$ qubits.
There are therefore $O(q/k)$ base cases.
The recursion is a binary tree hence there are $O(q/k)$ shortened product-phase operations in Step~4.
By Theorem~\ref{thm:exact}, each base-case call uses
$O\left(k(\log k)^\alpha\right)$ gates.
Each shortened product phase uses $O(M(k))$ gates by Lemma~\ref{lem:product-phase}.
Together they use
\begin{equation*}
O\left(\frac{q}{k}\left(k(\log k)^\alpha+M(k)\right)\right)
=O\left(q(\log k)^\alpha\right)
\end{equation*}
gates.

We postpone the exchanges in Step~6, keeping track of which wire holds each qubit
and applying subsequent gates to those wires.
This choice of wires is made when constructing the circuit and requires no gates.
At the end, one permutation puts all output qubits in the required order using $O(q)$ swaps.

Each base-case call uses $O\left(k(\log k)^\alpha\right)$ auxiliary qubits
by Theorem~\ref{thm:exact}, and each product-phase operation uses $O(M(k))$
by Lemma~\ref{lem:product-phase}.
These operations run sequentially and return their auxiliary qubits to zero,
so they can reuse the same register $A$.
Thus $a=O\left(k(\log k)^\alpha\right)$ auxiliary qubits suffice,
and the circuit uses $q+a$ qubits in total.
Using $k=\Theta(L)$, the resulting bounds are
\begin{equation*}
O\left(q(\log L)^\alpha\right)\quad\text{gates},
\qquad
O\left(L(\log L)^\alpha\right)\quad\text{auxiliary qubits}.
\end{equation*}

If $q\leq2k$, there is only one $\mathsf{ExactQFT}$ call.
Its gate count and auxiliary space are both $O\left(q(\log q)^\alpha\right)$, where $q=O(L)$.

\subsection{Replacing the additional rotations}\label{sec:hand-approx-rotations}

Recall that $\mathsf{ExactQFT}$ uses the circuit $T=UR_0U^\dagger D U$
and its inverse from Section~\ref{sec:proof-exact}.
These make $O(1)$ calls to the circuit $U$ and its inverse,
each of which contains one additional rotation or its inverse.
Thus $\mathsf{ExactQFT}$ uses $B(\theta_t)$ and its inverse only $O(1)$ times.
The following lemma bounds the resulting error on the input and auxiliary registers.

\begin{lemma}\label{lem:hand-rotation-replacement}
Let $t\geq64$ and let $X_t$ be the unitary implemented by the complete
$\mathsf{ExactQFT}$ circuit on its input and auxiliary registers.
Replacing its additional rotations gives a unitary $\widehat X_t$ using only
Hadamard gates, controlled-NOT gates, and the phase gates in
Equation~\ref{eq:phase-gates}, with
\begin{equation}\label{eq:hand-rotation-error}
\norm{\widehat X_t-X_t}\leq12\pi2^{-k}
\end{equation}
in operator norm.
Each rotation replacement uses a constant number of gates and no additional auxiliary qubits.
\end{lemma}

\begin{proof}
The idea is to express the rotation using a phase gate and then round its angle.
Set $S=P\left(\pi/2\right)$ and $J=SH$.
Direct multiplication gives
\begin{equation*}
JP\left(2\theta_t\right)J^\dagger=e^{i\theta_t}B\left(\theta_t\right).
\end{equation*}
Choose an integer $a_t$ such that
\begin{equation*}
\varphi_t=\frac{2\pi a_t}{2^k},
\qquad
\left|\varphi_t-2\theta_t\right|\leq2\pi2^{-k},
\end{equation*}
and put $C_t=JP\left(\varphi_t\right)J^\dagger$.
Replace $B\left(\theta_t\right)$ by $C_t$ and its inverse by $C_t^\dagger$.
These use only a constant number of allowed gates, since $S$ and $S^\dagger$
are themselves phase gates of the form in Equation~\ref{eq:phase-gates}.
Moreover,
\begin{equation*}
\norm{C_t-e^{i\theta_t}B\left(\theta_t\right)}
=\left|e^{i\varphi_t}-e^{2i\theta_t}\right|
\leq2\pi2^{-k}.
\end{equation*}
The adjoints satisfy the same bound.

The scalar phases cause no error in the complete circuit. Replacing these six gates one at a time by $C_t$ or $C_t^\dagger$ gives total
operator-norm error at most $6\cdot2\pi2^{-k}$ by the triangle inequality.
\end{proof}

Apply Lemma~\ref{lem:hand-rotation-replacement} to every base-case call in $U_0$,
and denote the resulting circuit by $\widehat U_0$.

\paragraph{Error and resource bounds after replacement.}
The auxiliary register $A$ may no longer return exactly to zero after a base-case call.
We pass it to the next operation without measuring or resetting it,
and include it throughout the error comparison.
Lemma~\ref{lem:hand-rotation-replacement} bounds the change in each complete call
by $12\pi2^{-k}$ in operator norm, including its auxiliary registers.
This bound applies even when those registers contain errors or are entangled
with other registers.
There are at most $q$ base-case calls, so replacing them one at a time gives
\begin{equation*}
\norm{\widehat U_0-U_0}\leq12\pi q2^{-k}.
\end{equation*}
Combining this with Equation~\ref{eq:hand-shortening-error} gives
\begin{equation}\label{eq:hand-approx-total-error}
\norm{\widehat U_0\left(\ket{\psi}\ket{0^a}\right)
 -(F_q\ket{\psi})\ket{0^a}}
\leq16\pi q2^{-k}\leq\varepsilon.
\end{equation}

\section{Open problems}

Our results leave several questions about the circuit complexity of the QFT. We mention some of them in this section.

Our exact construction uses an additional single-qubit rotation whose
angle depends on $q$.
Can the same gate bound be achieved using only Hadamard, controlled-NOT,
and the phase gates in Equation~\ref{eq:phase-gates}?

We do not establish uniformity of our circuit families in this work.
Assuming that the multiplication circuits can be generated in classical
polynomial time, can our QFT circuits also be generated efficiently?
It would suffice to show that the additional rotation angle, determined
by the average success probability of the approximate inverse QFT,
can be approximated to additive error $2^{-d}$ by a classical algorithm
running in time polynomial in $q$ and $d$.

Finally, can the exact QFT on $q$ qubits be implemented using $O(q)$
one- and two-qubit gates?
This question was raised by Aaronson~\cite{Aaronson2025QFT}.
The analogous question is also natural for approximate QFTs:
for a fixed constant $0<\varepsilon<1/2$, can
$\varepsilon$-approximate QFT be computed by a circuit of size $O(q)$?

\paragraph{AI statement.} We used GPT-6 Astra, GPT-6.1 Sol, and Claude Opus 5.5 to explore proof
strategies, search for related literature, and revise the manuscript.
The main proof ideas were generated by these systems.
The authors examined these ideas, identified and corrected errors, and
significantly simplified the arguments and their presentation.
The authors take full responsibility for the correctness and content
of the paper.

\bibliography{reference-v2}

\end{document}

%% file: figures/ach-recursion.tex
\begin{tikzpicture}[x=1cm,y=1cm,font=\normalsize,
  block/.style={draw,fill=white,line width=0.7pt,align=center}]
\foreach \y in {-1.2,0,1.2}
  \draw[line width=0.6pt] (0,\y) -- (8.4,\y);
\node[anchor=east] at (-0.2,1.2) {Upper register};
\node[anchor=east] at (-0.2,0) {Lower register};
\node[anchor=east] at (-0.2,-1.2) {Workspace $\ket{0}$};
\node[anchor=west] at (8.6,-1.2) {$\ket{0}$};
\node[block,minimum width=1.75cm,minimum height=0.7cm]
  at (1.65,1.2) {Recursive\\QFT};
\node[block,minimum width=1.7cm,minimum height=3.1cm]
  at (4.2,0) {Product\\phase};
\node[block,minimum width=1.75cm,minimum height=0.7cm]
  at (6.75,0) {Recursive\\QFT};
\end{tikzpicture}